# Finding a targetable super-hub within the network of cancer cell persistency and adaptiveness: a clinician-scientist quantitative perspective for melanoma

Rastine Merat[1, 2*]

(December 2022)

[1]*Dermato-Oncology Unit, Division of Dermatology, University Hospital of Geneva, Switzerland*
[2]*Department of Pathology and Immunology, Faculty of Medicine, University of Geneva, Switzerland*

In this perspective article, a clinically inspired phenotype-driven experimental approach is put forward to address the challenge of the adaptive response of solid cancers to small-molecule targeted therapies. A list of conditions is derived, including an experimental quantitative assessment of cell plasticity and an information theory-based detection of *in vivo* dependencies, for the discovery of post-transcriptional druggable mechanisms capable of preventing at multiple levels the emergence of plastic dedifferentiated slow-proliferating cells. The approach is illustrated by the author's own work in the example case of the adaptive response of *BRAFV600*-melanoma to BRAF inhibition. A bench-to-bedside and back to bench effort leads to a therapeutic strategy in which the inhibition of the baseline activity of the interferon-γ-activated inhibitor of translation (GAIT) complex, incriminated in the expression insufficiency of the RNA-binding protein HuR in a minority of cells, results in the suppression of the plastic, intermittently slow-proliferating cells involved in the adaptive response. A similar approach is recommended for the validation of other classes of mechanisms that we seek to modulate to overcome this complex challenge of modern cancer therapy.

## I. Introduction

The importance of the adaptive response within the mechanisms of resistance to solid cancer therapies, became obvious to clinician-scientists more than a decade ago upon the observation that solid metastatic tumors, such as seen in *BRAFV600*-mutated metastatic melanoma disease, disappeared in a spectacular manner under the effect of BRAF inhibitors but relapsed precisely on the same clinically or radiologically defined initial affected sites in a majority of patients. Indeed, contrary to what it was initially suggested, these multiple simultaneous and similar regrowth dynamics in a given patient could not be attributed to the occurrence of synchronous independent clonal mutations at each affected site [1]. Within the same paradigmatic example of *BRAFV600*-mutated melanoma disease, the importance of the adaptive response for patient outcomes became even more obvious as therapeutic trials demonstrated that 70% of the 20% of patients who do not show any clinical or radiological sign of immediate adaptive response to BRAF/MEK inhibitors, i.e., have an initial complete response, have durable responses [2]. In this minority of unpredictable complete responders, the percentage of long responders was in fact even higher than those obtained with combined checkpoint inhibitor-based immunotherapies [3]. Importantly, although combinations of checkpoint inhibitors, recommended nowadays as first-line therapies in metastatic melanoma, have demonstrated their overall superiority to small-molecule targeted therapies in terms of durability of response, they frequently induce immune system dysregulations with irreversible consequences. Therefore, subject to a better understanding of the mechanisms of the adaptive response and an effective prevention of it, small-molecule targeted therapies could come back to the forefront of cancer therapeutics which, while being effective, would not induce any permanent damage.

Here, it is important to clarify that the terminology "adaptive response" is preferred to "Lamarckian adaptation", increasingly used in the literature. Indeed, a "Neo-Lamarckian adaptation" occurring under the effect of environmental and therapeutic selection pressure, implies, similarly to the opposed genetic mutation-based "Neo-Darwinian adaptation", that the molecular modifications are transmissible from one cell generation to the next. Such transmissibility is certainly conceivable if we refer only to mechanisms involving epigenetic DNA modifications, but the mechanisms involved in adaptive response also call upon all sorts of other post-transcriptional and post-translational molecular events, and even metabolic changes, which are for the most part reversible and not necessarily transmissible. As further emphasized in this perspective, reversibility is even a necessary condition to be fulfilled for most mechanisms incriminated in adaptive response, without which many phenotypic observations, including a restored response for patients who had a previous therapeutic failure, would remain unexplained.

## II. A common slow-proliferating phenotype for variously defined cell-states involved in the adaptive response

Established whether clinically or from a cancer cell biologist perspective, a wide range of classifications are used to phenotypically characterize cancer cell subpopulations involved in the immediate or long-term adaptive response to micro-environmental changes and therapeutic exposure [4-10]. Dormant and persistent cancer cells may for instance be, but not necessarily, clinically distinguished based on being either early disseminated cells with long term latency, or alternatively residual surviving drug-tolerant cells, but both can be the constituents of a non-responsive or undetectable residual disease from which disease progression occurs. From the cancer cell biologist point of view, an adaptive mechanism does not necessarily occur from a subpopulation of cells, and many experimentally discovered mechanisms of adaptive response were attributed, under experimental conditions, to the bulk of malignant cells analyzed *ex* or *in vivo* [11-14]. However, as seen in *BRAFV600*-melanoma disease, a quasi-complete regression of metastatic tumor masses under the effect of small-molecule targeted therapies, and their subsequent resurgence in the same sites from undetectable residual cells, implies that the adaptive response emerges from a minority of cells. Although not formally demonstrated, the involvement of a minority of cells in the adaptive response likely applies also to the often-distinguished concept of tumor mass dormancy in which, through simultaneous growth and death, tumors cells are renewed under environmental selective conditions in a residual disease which remains undetectable [15]. Therefore, large efforts to understand the mechanisms of adaptive response have lately been justifiably focused on characterizing subpopulations of cells involved in the adaptive response.

---
[*] rastine.merat@unige.ch

Overall, apart from their ability to escape from the immune system and to adapt to their microenvironment, the most common denominator of these cells is recognized to be their slow-proliferating phenotype (alternatively, many authors prefer the term "slow-cycling cells" / phenotype). This phenotype is considered to cover a spectrum, a reversible senescent-like phenotype being the extreme quiescent state within this spectrum [9]. However, recently, an even more flexible model has been put forward, based on the observation that tolerant non-proliferating cells can occur stochastically from any proliferating cells rather than from a specific cell subpopulation upon exposure to environmental changes including drugs [16]. This observation has led the authors to describe a state of "diapause". Importantly, this model does not contradict one in which only a minority of cells are, at any given time, commited to the adaptive response; however, it does imply that the deceleration mechanisms are reversible, "sliding" and interchangeable among cells. In an effort to identify reversible mechanisms, some investigators have pushed to consider persistent slow-proliferating cells *per se* and elucidate the mechanisms of cell deceleration in these cells independently from any predefined cell state, i.e., senescence, diapause or even stemness [17-18]. Indeed, the other extensively studied cell state consistent with persistency or dormancy and associated with adaptive response, is the cancer stem-cell dedifferentiated phenotype [19]. Stemness requires the occurrence of asymmetric division and a cell hierarchy filiation to slow-proliferating cells, which are intimately affiliated to drug-induced senescent cells [20,21]. Here again, a spectrum of dedifferentiation occurs as in the example of melanoma cells, which may lose the expression of the microphthalmia-associated transcription factor MITF [22,23] and the associated melanocytic lineage differentiation markers, or even transdifferentiate into a neural crest stem cell-like state [24]. Overall, our awareness of the reversibility of the slow-proliferating and dedifferentiated phenotypes of the minority of cells involved at any time point in the adaptive response, has been of paramount importance to envision the molecular mechanisms we seek to target to prevent the adaptive resistance to small-molecule targeted therapies.

Generally, different classes of adaptive mechanisms have been explored as targets to either eliminate/prevent the pool of slow-proliferating cells or to reengage them in fast cell-proliferation. In that case, as discussed below, the targeted mechanisms should operate on a clearly distinguishable minor cell subpopulation to avoid any increase of the tumor mass. Here, representative examples for each class are listed.

**(i) Metabolic mechanisms:** A switch from an inefficient high-glucose consumption state in which the tricarboxylic acid cycle is excessively used in fast-proliferating cells (Warburg effect), to a low-glucose consumption state with partial restored mitochondrial oxidative phosphorylation in slow-proliferating cells has been documented [18]. The blockage of this switch by inhibiting the mitochondrial respiratory chain has been proposed as a potential strategy to prevent the emergence of slow-proliferating cells. Similar strategies are envisioned for other metabolic changes observed in slow-proliferating cells related to the use of alternative nutrients as increased autophagy [25] and/or the excessive use of fatty acid oxidation [26].

**(ii) Epigenetic mechanisms:** Epigenetic reprogramming has been shown to act as a control parameter of the state of proliferation. The phenotypic plasticity of slow-proliferating senescent-like cells depends at the transcriptional level on chromatin remodeling, which is not necessarily detectable by any change in senescence-associated heterochromatic foci (SAHFs), but occurs as discrete remodeling of the facultative heterochromatin. Such discrete remodeling depends on the changes in histone marks, as for example H3K4me3, which are themselves dependent on histone demethylases activity. Therefore, direct inhibition of the involved demethylase [27] or inhibition of the activating upstream signaling pathway, as the IGF-1R signaling, represent potential strategies to eliminate slow-proliferating cells [17,18].

**(iii) Transcriptional mechanisms:** By using a TEAD inhibitor to inhibit the transcriptional YAP-/TEAD/SLUG complex and further restore *BMF* expression, a team was able, upon enhanced EGFR/MEK inhibition-induced apoptosis, to deplete dormant cells in a model of *EGFR*-mutant non-small cell lung cancer [28]. Another representative example of this class is the targeted inhibition of the dual-specificity tyrosine phosphorylation-regulated kinase 1A (DYRK1A) involved in the assembly of the DP, RB-like, E2F4 and MuvB (DREAM) complex which represses cell cycle-dependent genes during quiescence and senescence [29].

In this perspective, I propose an alternative class of RNA-binding protein (RBP)-mediated targetable **(iv) post-transcriptional mechanisms** that aim to avoid the emergence of dedifferentiated slow-proliferating cells by interfering at multiple levels within the network of the adaptive response. The initial motivation to explore targetable post-transcriptional mechanisms for this important therapeutic matter stems from the results of a number of theoretical *in silico* studies showing that post-transcriptional fluctuations, rather than transcriptional fluctuations, best coordinate genes and control macro-heterogeneity, i.e., the number of possible trajectories in gene expression, including bimodalities, at least on long-enough time scales [30-32].

### III. Conditions to incriminate mechanisms involved in the dedifferentiated slow-proliferating phenotype

The clinically and experimentally based understanding of the adaptive response described above leads to the below-listed qualitative and quantitative conditions that should be met by the investigated mechanism. The fulfillment of some of these conditions at bench in predictable and reproducible homogenous cell lines / xenograft models previously well-characterized is mandatory. Indeed, unlike many experimental settings where heterogeneity is sought to mimic *in vivo* tumor heterogeneity, here, a "reductionist clean" homogeneous cell population is a required starting condition.

**1-** It should be an actionable mechanism that controls the occurrence of slow-proliferating/senescent-like cells. A large body of independently conducted research should support its involvement in the control of cell proliferation. Its mechanistic link with the expression control of genes involved in cell-cycle regulation, should be clearly established in the literature.

**2-** The mechanism should operate in a minority of cells at any time point.

**3-** In order to avoid any further proliferation in the remaining dominant proliferating cell population, the mechanism should operate, at least in experimental conditions (difficult to observe *in vivo*), in one mode of a bimodal distribution within only the minority of slow-proliferating cells (as opposed to a gradual distribution in the whole cell population), i.e., its modulation in the minority of slow-proliferating cells should not affect the remaining dominant cell population.

**4-** The mechanism should be reversible.

**5-** The mechanism should be tested *per se* as a reversible mechanism; indeed, since the slow-proliferation state can reversibly emerge stochastically from any proliferating cells, the effects of the experimentally modulated mechanism should be monitored under reversible conditions within the same cell population. This is to confirm that the interchangeability among cells of the tested mechanisms has the expected consequences on the adaptive response. In other words, reversibility should be considered as part of the mechanism.

**6-** The plasticity of the proliferation state (slow *versus* fast) should be detectable within the subpopulation of cells in which the mechanism is at play, in other words its involvement in the simplest form of hierarchy or stemness should be tested. This means testing if the slow-proliferating state can induce a fast-proliferating adapted state (asymmetry of filiation even though the reverse filiation operates as well in stochastic reversibility from fast to slow-proliferating cells). Since stochastic reversibility and interchangeability among cells imply that no barcoding can be used for long-enough cell-tracking adaptive experiments, I propose a static cytometry-based expression analysis of a large enough panel of markers to clearly define cell subpopulations (more feasible with mass cytometry) coupled with viSNE/SPADE analysis [33, 34], to assess if the distribution of the two states is scale-invariant, in other words if it has a "fractal" distribution. Quantitatively, any more narrowly defined subpopulation of cells in which the mechanism operates within a more largely defined ensemble/cluster of such cells should carry, similarly to the larger ensemble, roughly the same probability distribution for the two states over a time period in an adaptive experiment. Even if not practically performed, when iterating this process, even a single cell within this subpopulation should have the same probability distribution for the two states over a time period. This ensures that the incriminated subpopulation carries the plasticity of the more largely defined cell population. Importantly, unsupervised clustering should be mathematically defined among many combinations of varying viSNE and SPADE parametrizations (Fig. 1) [35,36].

**7-** The mechanism should operate as a major controlling hub within the signaling network known to control cell proliferation e.g., MAPK and/or PI3K-AKT, which are inactivated in slow-proliferating quiescent/dormant cells [37,38]. Importantly, the *in vivo* dependency between the incriminated mechanism and the signaling activity of these pathways should be demonstrated. Considering the nonlinear and time-shifted dependencies of the involved mechanistic factors that are assumed to operate in a minority of cells (condition 2) and to be reversible (condition 4), I propose an information theory approach in which a copula/kernel density estimator (KDE) is used as a nonparametric method to estimate probability densities for the quantification of mutual information (MI) [39]. We established that this method of measuring dependencies is far more sensitive than average co-expression or even single-cell-based co-expression correlation coefficients. Moreover, we validated it regarding various bias [40], including the size of the probability distribution using mouse xenografts upon tumor growth, and the sample size using patient's metastatic samples to ensure tumor heterogeneity [41]. This approach is also particularly suited for the translational analytic pipeline that would result from this type of experimental research. Indeed, average expression measurement of MAPK or PI3K/AKT activation markers often fails to predict response to therapy, as shown in patients with BRAFi-resisting metastatic melanoma disease [42]. This is indeed not surprising if we consider that the phenotypic response of any tumor is not determined by the majority of proliferating cells but rather by the proportion of the difficult-to-detect, highly diluted minority of plastic slow-proliferating cells.

**8-** The mechanism should be clearly involved in the control of additional mechanisms involved in dedifferentiation as well as epithelial mesenchymal transition (EMT). In *BRAFV600*-melanoma, examples of such mechanisms include: the mild dedifferentiated drug-induced MITF$^{Low}$ state associated with an EMT gene expression signature and resistance to MAPK inhibitors [23], and the EMT-like signaling associated with slow proliferation, which occurs in EGFR-expressing cells chronically exposed to a BRAF inhibitor (BRAFi) [43]. Moreover, overexpression of other receptor tyrosine kinases (RTKs) as IGF-1R, MET or FGFR3 has also been associated with EMT in various malignant contexts [44-46]. As further explained below, these down- or upregulations may occur at the post-transcriptional level. Additional mechanisms of EMT involving change in isoform usage, which are in essence post-transcriptional events, include the expression of *CD44* isoforms harboring the variable exons (CD44s), instead of the isoforms devoid of them, which is mandatory for EMT [47, 48]. Here, again, the *in vivo* dependencies between the main targeted mechanism and the additional mechanisms incriminated in dedifferentiation or EMT, should be verified as in condition 7, using the proposed information theory approach.

**9-** The mechanism should be modulated by some of the already identified environmental triggers of slow-proliferation and or senescence, as for example the interferon gamma (IFN γ) or the TNF signaling pathways [49,50].

**10-** The mechanism should be drug targetable i.e., an involved druggable enzyme should be actionable in the mechanism.

### IV. The human antigen R (HuR): a perfect candidate wrongfully accused?

Among the subfamily of regulatory mRNA-binding proteins, HuR (Embryonic Lethal Abnormal Vision 1/ELAVL1) has naturally drawn our attention as one of the rare post-transcriptional *trans*-regulators of both cell proliferation and differentiation. HuR is a ubiquitously ex-pressed RBP that, similarly to other regulatory RBPs, has the ability to rapidly control the fate of mRNAs in response to environmental changes including through its nucleo-cytoplasmic shuttling which is post-translationally determined [51, 52]. HuR was originally discovered in 1985, along with its paralogs by screening a human cerebellum cDNA expression library with sera from patients affected by an autoimmune opsoclonus-myoclonus paraneoplastic syndrome [53]. The human HuR gene was subsequently cloned in 1996 [54]. Based on extensive studies during the last two decades of its repertoire which covers a large part of the transcriptome, HuR is now considered to be a major hub within the post-transcriptional network of regulatory RBPs [55-59]. This centrality suggests that despite being tightly regulated, slight variations of HuR expression could be enough to induce large perturbations within its direct or indirect repertoire and consequently determine the all-or-none switching in individual cells within the post-transcriptional landscape of possible attractor states [60].

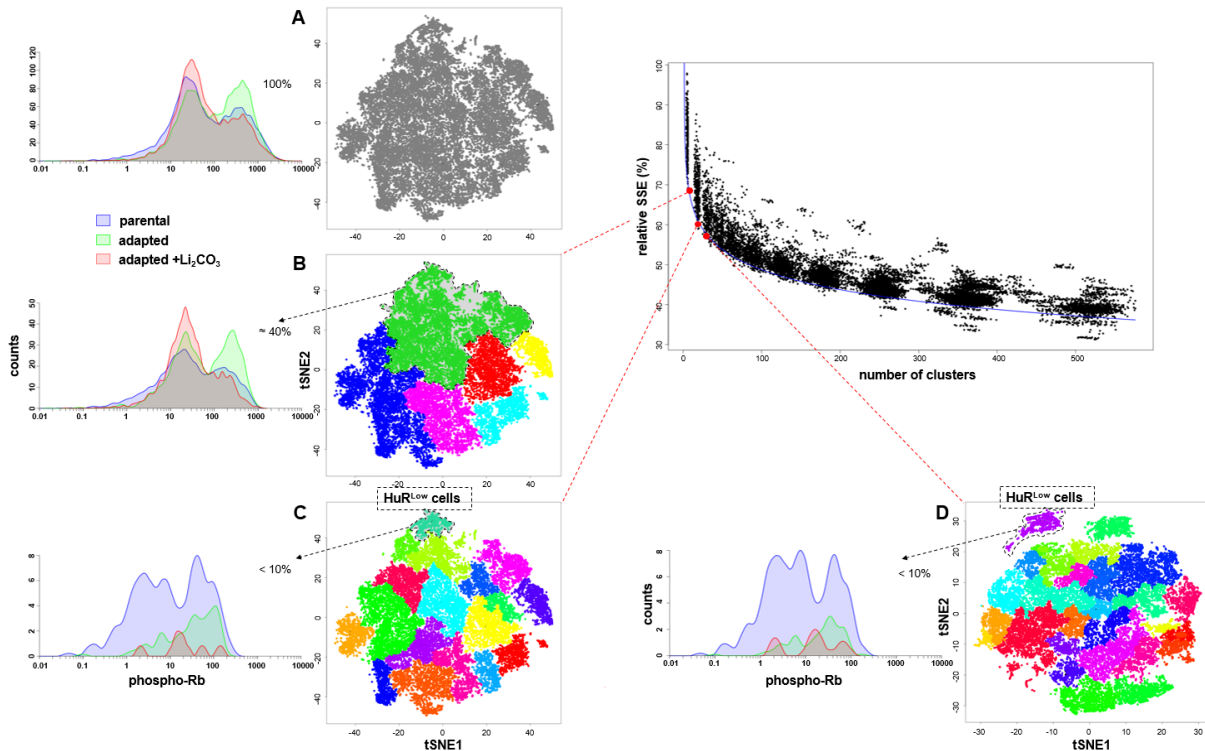

**Fig 1**. **Plasticity of the state of proliferation defined by its static scale-invariance in the adaptive response.** In the example case of our own work in which an HuR$^{Low}$ state is incriminated in the adaptive response, A375 melanoma cells were exposed to an adaptive regimen i.e., incremental doses of vemurafenib with or without concomitant exposure to a drug (lithium salts) that reduces the proportion of HuR$^{Low}$ cells; cell subpopulations were then defined in the parental and adapted +/- lithium salts-treated cells based on the expression of 13 markers using mass cytometry. Regardless of the clustering scale used: (i) whole cell population (A), (ii) a largely defined ensemble/cluster of cells comprising the HuR$^{Low}$ cells (B), (iii) a more narrowly defined sub-ensemble/cluster of HuR lowest expressing cells (C), the same roughly defined probability distribution is observed for slow (phospho-Rb low) and fast (phospho-RB high) -proliferating cells. Importantly, this roughly equiprobable distribution was only detected in the largely and more narrowly respectively defined HuR$^{Low}$ clusters and not in any of the other defined clusters in B and C. Moreover, in the narrowly defined HuR$^{Low}$ cluster in C, this probability distribution is only observable in the parental cells, while it is "transferred" to the more largely defined subpopulation and the whole cell population in adapted cells (unless treated with lithium salts). Mathematically, an unsupervised approach is used to choose the SPADE-defined clustering scales in B and C, within the same viSNE map, in an optimal area of maximum values of a curvature function (knee approach) in the graph of a large number of possibilities in which the number of clusters is plotted as a function of their relative error sum of squares (SSE). A different viSNE map, also chosen within the optimal area, is shown in which the HuR lowest expressing cluster is even better individualized (D). In this cluster again, a roughly equivalent (to C) probability distribution is observed for phospho-Rb expression. The detection of the incriminated cell subpopulation characterized by a scale-invariant plasticity for cell proliferation, is therefore not "clustering-dependent".

HuR has the ability to control key post-transcriptional processing steps for some of its mRNA targets such as splicing, nuclear export [51,61,62], stability [63,64] and translation [65, 66]. Although HuR was initially characterized as having stabilizing effects on many mRNAs and by its ability to increase their translation, its antagonistic effects on a subset of its repertoire have also been demonstrated [67,68]. These differential effects are thought to be primarily due to the interaction between HuR and numerous microRNAs that either compete or cooperate for their resulting effects on mRNA targets [69]. Interestingly, HuR has the ability to attenuate microRNA-mediated repression by inducing, in its oligomeric form, dissociation of the miRISC complex from at least some of its targets [70, 71]. Regarding HuR antagonistic effects, they may also operate via cap-dependent or IRES-mediated inhibition of translation initiation of its mRNA targets [72, 73].

Generally, deducing the phenotypic consequences of an RBP expression modulation based on the known function of any of its demonstrated targets is not reliable, especially in transformed cells, considering that the post-transcriptional expression of potential targets has been shown to be primarily differentially regulated rather than absolutely determined by oncogenic signaling [74]. In this system, the phenotypic consequences of *trans*-regulation are more dependent on a "relative to other targets" rather than an absolute effect on any target and therefore are most likely affected by the baseline expression repertoire of cells. This has been shown in malignant cells for some RBPs that differentially affect the fate of their mRNA targets [75, 76]. In spite of this complexity, HuR involvement in the control of cell proliferation is widely established and overall operates mechanistically through a post-transcriptional expression-synchronizing effect on cell cycle regulatory genes/targets as cyclin A2, cyclin B1, cyclin D1, cyclin E, $p16^{INK4}$, $p19^{ARF}$, p21, p27, WEE1, p53 and Mdm2 [64,65,72,77-86].

Importantly, loss of HuR expression observed in fibroblasts has been linked causatively to replicative senescence, and mechanistically to reduced expression of genes involved in cell cycle progression as cyclin A and B1, and to increased expression of cell cycle inhibitory genes as $p16^{INK4}$, $p19^{ARF}$ [80,81,87] and even p27. Regarding the latter, although diametrically opposite effects of HuR have been reported [72,82,83], the phenotypic effect of insufficient or excessive accumulation of inhibitors as p27 (or p21) might be counter-intuitive, considering that cell cycle progression or inhibition is not only determined by their inhibitory activity on cyclin-CDK complexes in late G1 through M phase but also by their stimulatory sequestration by cyclin D-CDK4/6 complexes in the early and mid-G1 phase [88, 89]. Again, in this system, rather than an absolute post-transcriptional effect of HuR, it is more an effect on p27 (or p21) expression relative to that exerted on various cyclins that will ultimately determine cell cycle progression or inhibition. This also illustrates how HuR robustly coordinates the cell cycle across multiple redundant targets/mechanisms.

Senescence associated with HuR insufficiency also relates to senescence-associated secretory phenotype (SASP) and particularly the secretion of IL-6 [90], which was also observed in an *in vivo* model of myeloid lineage HuR insufficiency [91]. In this model, the HuR insufficient-induced pro-inflammatory response and resulting colitis was also associated with TNF, CCL2 and INF-γ excess production, a profile resembling to SASP.

Another important phenotypic effect of HuR is its ability, as its paralogs, to force differentiation in various differentiated cellular contexts including melanocytes [92-94]. In HuR overexpressing normal melanocytes for example, the translation of melanocytic markers as dopachrome tautomerase is increased. An increase in the expression of the differentiation-inducing transcription factor MITF has also been shown in HuR-overexpressing *BRAF*-wild type melanoma cells [95]. Therefore, although the molecular mechanisms involved in differentiation are often opposed to those involved in proliferation, HuR *trans*-regulatory effects seem to operate by a synchronizing effect on the expression of either cell cycle regulatory genes in proliferating cells, or differentiation-associated genes in differentiated cells.

Regarding the potential involvement of HuR in change in cell identity, some observations support the involvement of its insufficient expression in EMT. First, our results indicate that EGFR and MET are upregulated upon HuR knockdown in *BRAFV600*-melanoma cells chronically exposed to a BRAFi [35]. Although we did not elucidate the underlying mechanism of this upregulation, other genome-wide studies have identified both EGFR and MET as HuR potential direct targets [56,57]. IGF-1R expression is also markedly increased upon HuR knockdown, and conversely HuR overexpression represses IGF-1R expression by differentially repressing cap-dependent and IRES-mediated translation of IGF-1R mRNA in the generic context of Hela cells [73]. Another recently elucidated mechanism linking HuR insufficiency to EMT, is its involvement in determining *CD44* splicing variants usage, HuR insufficient expression being associated with the EMT inducing isoforms CD44s [62]. Finally, non-canonical Wnt signaling and more particularly Wnt-5a which has been implicated both in EMT and dormancy and even melanoma adaptive response [14,96,97], is negatively regulated by HuR at the post-transcriptional level [98]; HuR insufficient expression could therefore be involved in Wnt-5a upregulation in EMT.

In contrast to what is observed in senescent cells, HuR expression is increased in proliferating cells within regenerating tissues and proliferative tumors [99,100]. However, at least according to our observations, no matter them being benign or malignant (Fig. 2). This is in fact the case for many genes involved in cell cycle regulation, whether being tumor suppressors or pro-tumorigenic [102-104]. Nevertheless, an important legitimate question arising from these observations is whether HuR should be considered a pro-tumorigenic RBP. Overall, the available *in vivo* experimentally obtained information does not support this hypothesis since – although as expected, HuR knockdown hinders cell proliferation and tumor formation – overexpression experiments have not demonstrated [91] or failed in demonstrating [105] tumor development. In fact, HuR overexpression protected mice not only from pathologic inflammation but also from colorectal carcinogenesis. Regarding the expression analysis of HuR as a prognostic marker in cancer, studies in which its expression was analyzed as a predictor of histological grading rather than clinical outcome, have shown that HuR increased expression or cytoplasmic detection is associated with a higher tumor grade [100], higher grade being itself often associated with a higher proliferation index. However, in the same malignant disease, when analyzed as a predictor of metastatic events, HuR higher expression was associated with a favorable outcome [106]. Here, I would like to emphasize that although grading is used routinely to predict disease outcome, it often fails to do so, the phenotype of any malignancy being only definable by its effective clinical outcome. As an example, in the case of melanoma, tumor thickness (Breslow index) is used as a marker to predict metastatic outcome and thicker tumors, having often also higher mitotic activity, have a worse outcome. Nevertheless, approximately half of these thick "high-risk" tumors fortunately do not induce metastatic disease. We have ourselves explored this question in melanoma primary tumors and observed that, rather than the expression level of HuR, it is its higher level of expression heterogeneity which is associated with metastatic dissemination [107].

### V. HuR insufficiency as a targetable mechanism: a "bench-to-bedside and back to bench" approach

As argued, HuR is a key player in the post-transcriptional expression regulation of genes controlling the cell-cycle, and being indispensable for cell proliferation, its insufficient expression does provoke senescence. Moreover, as discussed, HuR has the ability to force differentiation and its insufficient expression is potentially involved in the expression of EMT-inducing surface receptors. Considering that condition 1 listed above was met (and even parts of condition 8), we initiated a "bench-to-bedside and back to bench" approach to explore HuR potential involvement in the adaptive response.

In a series of initial *ex vivo* experiments, we observed that HuR transient overexpression could attenuate the immediate paradoxical proliferation of *BRAFV600*-melanoma cells to BRAF inhibition [108]. An attenuation of the long-term adaptive proliferation was later also confirmed *ex vivo* upon HuR stable overexpression [109]. Importantly, we used in the transient overexpression experiments a single-cell mass cytometry approach that clearly showed that the adaptive response/proliferation does emerge from a minority of cells, even in a highly homogeneous cell line having a high propensity for adaptive response. Upon these initial observations, we reasoned that if the subpopulation of cells involved in the adaptive response, including to microenvironment changes, was a slow-proliferating cell subpopulation with some cells having potentially an intermittent lower expression of HuR, a more heterogeneous expression of HuR should be detectable in primary melanoma tumors more likely to metastasize. This hypothesis was tested with an automated immunohistochemistry-based and descriptive statistics-based analysis of HuR expression in melanoma primary tumors. By quantifying the dimensionless fourth moment of HuR expression distribution, which is the kurtosis (HuR K), we could show that HuR expression is much more homogeneous (HuR K significantly higher) in the non-metastatic group of tumors comparatively to tumors associated with micro- or macrometastasis at two-year follow-up. Interestingly, based on comparison of the area under the curves (AUC) of the receiver operating characteristic, HuR K appeared even as a more robust marker of metastasis than the Breslow thickness [107]. Obviously, to be applicable for patient management, these results would need to be confirmed on a large cohort of patients, nevertheless, they encouraged us to pursue our bench investigations.

In an effort to better characterize the available *ex vivo* cell-line tools needed for our investigations, we analyzed the variability of HuR expression distribution among different *BRAFV600*-melanoma cell lines and noticed that in the context of such homogeneous cell (-line) populations, HuR lower average expression was predictive of a higher propensity for adaptive response to BRAF inhibition [35]. "Back-to-bedside", we observed a strong trend (even if not statistically significant, due to the size of our cohort) toward an increased radiologically-assessed response to MAPK inhibition therapy in *BRAFV600*-metastatic tumors containing a lower proportion of $HuR^{Low}$ cells (lower Z-score < -2SD) [35]. At bench again, we compared HuR expression distribution in various *BRAFV600*-melanoma cell-lines, between the parental cells and their adapted-to-BRAF-inhibition counterpart and observed that the proportion of $HuR^{Low}$ cells tended to increase in adapted cells [35]. Moreover, in cell lines having a high propensity for adaptive response, HuR expression distribution would reproducibly adopt a bimodal distribution at least in the adapted cells (Fig. 3). This bimodality was highly suggesting that a mechanism affecting HuR expression operates at any given time only in a cell subpopulation.

Subsequently, we were able to generate, depending on the cell line, either a stable or a reversible knockdown of HuR. Although we did not elucidate the mechanism of the reversibility of HuR knockdown in the latter, a bimodal reversible $HuR^{Low}/HuR^{High}$ distribution was maintained in these cells for a time long enough to confirm *ex vivo* their enhanced propensity for adaptive response/proliferation to BRAF inhibition. Eventually, a complete and irreversible knockdown of HuR occurred even in these cells, allowing them to survive even higher concentrations of a BRAFi, but rendering them senescent, as predicted by other studies, and incapable of adaptive proliferation. At this stage, conditions 2, 3, 4 and 5 listed above were fulfilled but further mechanistic studies (see below) will demonstrate that the bimodality is the signature of a post-transcriptional mechanism of HuR insufficient expression. To further verify if condition 6 was also met, we performed a simultaneous mass cytometry analysis of 13 markers to clearly define various cell-subpopulations in parental and adapted cells, and observed that the paradoxical adapted cell proliferation originated from the subpopulation of cells carrying a $HuR^{Low}$ component. Importantly, when increasing the number of clusters while ensuring the use of optimized t-SNE embeddings and subsequent SPADE-defined number of clusters, both largely defined and more narrowly defined ensemble/cluster of cells carrying the $HuR^{Low}$ cell component had the same roughly defined probability for slow (phospho-Rb low) and fast (phospho-RB high) -proliferating states over a time period (Fig. 1) [35].

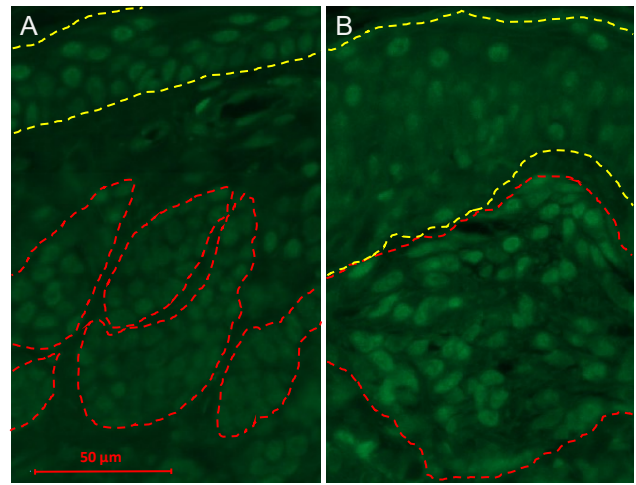

**Fig 2**. **Immunofluorescence-based HuR staining in naevi.** Typical HuR signal intensity in two types of benign naevi, one non-proliferative (Miescher dermal naevus, A), one proliferative (Spitz naevus, B). The HuR nuclear and cytoplasmic signals are more intense in the nests (within dashed-red lines) of the proliferative benign Spitz naevus (made of nests of fusiform cells arranged in interlacing bundles). The signal is homogeneous in both lesions unlike what we [107] and others [101] have observed in malignant skin tumors. HuR signal intensity in the epidermis (upper layer, within dashed-yellow lines), which is here similar in both images, is used as a normalization factor. Note that the epidermis is thicker and hyperplasic in B, a common hallmark of Spitz nevi.

HuR expression has been shown to be inducible by lithium salts by other investigators, however, in our hands its expression was only slightly inducible in various *BRAFV600*-melanoma cells and not in HuR knockdown cells used as a control condition [35]. Nevertheless, we could demonstrate that even a slight expression induction of HuR could strongly attenuate the adaptive response to BRAF inhibition both in *ex vivo* and in *in vivo* mice-xenograft experiments. This effect was shown to be HuR-dependent *ex vivo*. However, to demonstrate that the effect of lithium salt therapy was also HuR-dependent *in vivo*, we used an automated immuno-histochemistry-based assessment of the expression of HuR, EGFR and phospho-ERK resistance markers in biopsied tissues obtained from xenografts tumors during BRAFi +/- lithium salt therapy. Dependencies were then quantified using estimates of mutual information [41]. Importantly, we had previously demonstrated in our *ex vivo* experiments, in the bulk cell population, the existence of dependencies toward

HuR for a large panel of markers including some involved in MAPK and AKT signaling. In these experiments, a loss of MAPK and AKT activity in HuR$^{Low}$ adapted melanoma cells was clearly observed and prevented partially under concomitant lithium salt therapy [35]. Moreover, the loss of MAPK signaling was completely prevented under HuR stable overexpression, including when monitored at the single-cell level in time-lapse experiments using an ERK-activated KTR biosensor [109]. *In vivo*, in the xenograft biopsied samples, we could also clearly demonstrate that lithium salt therapy suppresses the HuR/EGFR and the BRAFi-induced HuR/pERK dependencies. These *in vivo* dependencies were shown to operate mainly in the HuR$^{Low}$ cell subpopulation in a separate HuR knockdown xenograft model [41]. Overall, condition 7 and 8 were met.

At this point, although we had identified lithium salts as a candidate drug to prevent to some extent HuR insufficiency and attenuate the adaptive response of melanoma cells to MAPK inhibition, the mechanisms of HuR insufficient expression remained unknown. Nevertheless, our above-mentioned observations of bimodality in HuR expression distribution, which was particularly detectable in the adapted melanoma cells, indicated that BRAF inhibition adaptive regimen could be used as a perturbation strategy to elucidate the mechanism of HuR insufficiency occurring at steady state in melanoma cells. In the bulk-cell population, under such perturbation, we were able to detect a decrease in HuR mRNA level which we could link to an increase in its decay rate. Based on 3'RACE RT-qPCR analyses, this decrease was attributed to the most abundant polyadenylation site (PAS) 2 HuR mRNA variant and was not associated with any change in isoform usage previously reported [110]. We then reasoned that *cis*-regulatory mechanisms, involved in HuR PAS2 mRNA insufficiency in a minority of cells, would become more detectable on overexpressed transcripts, and generated a series of mCherry-tagged HuR full-length PAS2 (PAS1-inactivated) and PAS1 mRNA constructs for which we could again observe a bimodal cell expression distribution in *BRAFV600*-melanoma transfected cells. Remarkably, the bimodality was only observed for the PAS2 construct in parental cells and was more pronounced in adapted cells, contrary to the PAS1 construct for which the bimodality was only slightly detected in adapted cells. The following cell-sorting-based separation of the two subpopulations in the PAS2-transfected adapted cells, demonstrated that the bimodality was related to two states of stability of the exogenous transcripts [109]. We then reasoned that the bimodality was more likely related to a regulatory on-off state-mechanism which occurs when the average time between the binding and unbinding of a regulatory factor dominates the time scale of other regulatory processes [111]. Moreover, assuming a unique regulator/mechanism and since the bimodality could also slightly occur for the PAS1 variant, we considered that robust micro-RNA/mRNA interactions, which should have had a similar effect on common sequences to both variants, were less likely involved. On the other hand, protein/mRNA interactions were more likely involved since less deterministic.

As a strategy to identify HuR PAS2 mRNA interacting proteins potentially involved in HuR insufficient expression, we used an MS2-BioTRAP under native condition coupled to liquid chromatography mass spectrometry (LC-MS) to capture not only RNA cross-linked proteins, but also non cross-linked protein-protein complexes [109]. In this strategy, a cluster of MS2 stem loops inserted downstream of the 3'UTR and upstream of the polyadenylation signal sequence, are recognized with nanomolar affinity by the bacteriophage protein MS2 which is tagged with a signal sequence for *in vivo* biotinylation and stably co-expressed to enable purification of the complexes [112]. With this approach, we were able to identify constituents of the interferon-γ-activated inhibitor of translation (GAIT) complex, including the glutamyl-prolyl-tRNA synthetase (EPRS) and the ribosomal protein L13a (L13a) which, respectively directly or indirectly, through the GAIT complex, interact with mRNAs. Importantly, although the activation of the GAIT complex has not been associated with mRNA decay but only with inhibition of translation initiation [113-118], we were able to demonstrate through the following successive observations its involvement in HuR insufficiency:

**(i)** Full-length EPRS was shown to interact with L13a in *BRAFV600*-melanoma cells, this interaction being expected to occur only if the GAIT complex is active;

**(ii)** Upon multiple sequence alignments from distant mammals, and by focusing on minimum-free energy (MFE) and centroid-based structure-preserving covariations that could not be attributed to phylogeny, we were able to identify a GAIT-like motif within a highly conserved region of the 3'UTR located upstream of the U-rich upstream auxiliary elements (USEs) of HuR transcript PAS2;

**(iii)** Within this motif, the mutation of the distal stem-loop structure so that it could no longer occur, or the insertion of an adenine in the distal loop to mimic a covariation found in the corresponding sequence of distant mammals as the ring tail lemur, the Chinese pangolin, the naked mole-rat or the guinea pig, had a stabilizing effect on HuR PAS2 full transcript. It is noteworthy that some of these mammals are characterized as being long-lived cancer resistant species;

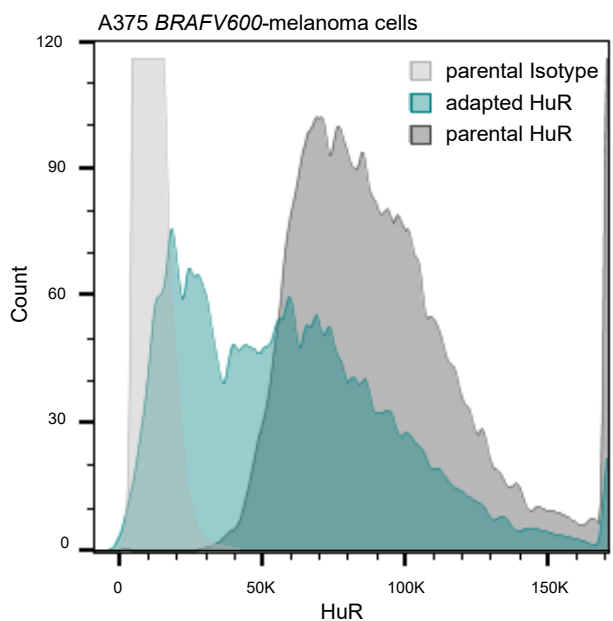

**Fig 3**. **Flow-cytometry analysis of HuR expression distribution in the adaptive response.** A375 *BRAFV600*-melanoma cells having a high propensity for adaptive response were exposed to incremental doses of vemurafenib up to 300 nM over 4 weeks to avoid any mortality and additionally treated with 1uM of vemurafenib for 6 h before being harvested. The bimodal distribution is detected, in this particular experiment, only in adapted cells in which the drug is less effective. Note that the proportion of HuR$^{Low}$ cells is particularly important here but may vary depending on the passage number of the parental cells. In similar experiments performed with different melanoma cell lines having various propensities for adaptive response, the dynamic of HuR expression distribution (between parental and adapted states) cannot be simply related to variations in the rate of proliferation of the whole cell population; indeed, the state of proliferation of HuR$^{Low}$ cells is at any given time undetermined (see Fig. 1). As argued in this perspective, the proportion of this subpopulation rather determines the plasticity of the whole cell population and its propensity to adapt to microenvironment changes including drug exposure (e.g., BRAFi treatment).

**(iv)** In L13a knockdown experiments and by using the above-mentioned mutant constructs, we were able to demonstrate that the dominant HuR PAS2 variant transcript is negatively regulated by L13a and that this regulation is dependent on the identified GAIT-like motif. Indeed, L13a knockdown increased the proportion of HuR$^{High}$ cells and attenuated HuR bimodal expression distribution for the exogenously expressed wild-type but not the mutated stem-loop- harboring constructs;

**(v)** In ribonucleoprotein immunoprecipitation (RIP) experiments, we could confirm that the EPRS-HuR PAS2 mRNA interaction is both L13a- and GAIT-like motif-dependent. Conversely, in MS2-BioTRAP experiments, we could confirm that HuR PAS2 mRNA indirect interaction with L13a protein is dependent on the GAIT-like motif.

These observations prompted us to test if the inhibition of the GAIT complex activation could prevent HuR endogenous expression insufficiency, and also the adaptive response of *BRAFV600*-melanoma cells to BRAF inhibition. In *ex vivo* adaptive regimen experiments, both L13a knockdown and the inhibition of the DAPK1/ZIPK axis, involved in the assembly/activation of the GAIT complex [119], with an oxazalone DAPK1/ZIPK inhibitor, attenuated the adaptive response/proliferation to BRAF inhibition and prevented the loss of MAPK signaling in adapted melanoma cells. Moreover, DAPK1/ZIPK inhibition was effective in stabilizing the HuR PAS2 transcript isoform and in reducing the proportion of HuR$^{Low}$ cells in the endogenous HuR bimodal expression distribution [109]. At this point, conditions 9 and 10 were also fulfilled, at least based on *ex vivo* observations. Indeed, *in vivo* experiments were hindered by the limited stability and the reactivity of oxazalone derivatives which are, to our knowledge, the most potent available dual DAPK1/ZIPK inhibitors [120].

Academic endeavors are currently being pursued to improve the potency of existing DAPK1/ZIPK inhibitors [121-128]; however, the discovery of more potent dual inhibitors of both kinases will more likely also rely on the efforts of pharmaceutical drug companies and on a more systematic exploration of the chemical space using predicative physics-based and machine learning methods [129]. Importantly, an efficient dual DAPK1/ZIPK inhibitor will potentially not only be useful to prevent the adaptive response to small-molecule targeted therapies of solid malignancies, as demonstrated by our own work for melanoma, but will likely have also other applications for other disease/conditions potentially associated with HuR insufficiency. These include conditions related to replicative senescence [90], chronic inflammation [91], diet-induced obesity [130] and non-alcoholic fatty liver disease [131]. Indeed, as remarkably demonstrated, mice bearing a constitutively inactive GAIT complex, due to phospho-deficient EPRS, are protected from most of these conditions and have a longer lifespan [132].

## VI. Conclusion

In this perspective, I formalize a list of validating experimental conditions, including quantitative constraints, as a general framework for the discovery of efficient targetable post-transcriptional mechanisms involved in the adaptive response of solid tumors to small-molecule targeted therapies. An important argument made here is that the quantitative fulfillment of these fundamental phenotype-driven properties of the adaptive response in well-characterized *ex vivo* or *in vivo* experimental reproducible cellular models, ensures much better the clinical relevance of mechanistic findings, than their validation in models that seek to mimic *in vivo* heterogeneity. These conditions have guided our own research to identify a targetable mechanism involved in the insufficient expression of HuR in a minority of cells in the context of the adaptive response of *BRAFV600*-melanoma to BRAF inhibition. However, validation of these conditions may prove necessary for any class of incriminated mechanism that we seek to modulate to effectively overcome the spontaneous or drug-induced emergence of plastic dedifferentiated slow-proliferating cells, be they epigenetic, transcriptional, post-transcriptional or even metabolic.

### Acknowledgments

I apologize to the researchers whose work was not cited because of space limitation. Some of the work discussed in this article was supported by the Ligue Genevoise contre le Cancer, the Fondation pour l'innovation sur le cancer et la biologie (Geneva), the Francis & Marie-France Minkoff Foundation and the Stiftung zur Krebsbekämpfung (Zurich).

### Declaration of interest

The author is inventor on a patent on the use of agents enhancing HuR/ELAV protein levels in the treatment of BRAF-mutated cancers.

---